\documentclass{article}

\usepackage{algorithm2e}
\usepackage{graphicx}
\usepackage{caption}
\usepackage{hyperref}  
\def\mun{\mathbf{\mu}}
\def\xn{\mathbf{x}}

\def\betan{\mathbf{\beta}}
\def\Sigman{\mathbf{\Sigma}}

\begin{document}

\title{Stage I non-small cell lung cancer stratification by using a model-based clustering algorithm with covariates.}
\author{Relvas C. and Fujita A. \\ Department of Computer Science \\ Institute of Mathematics and Statistics \\ University of São Paulo.}

\maketitle

\abstract{
Lung cancer is currently the leading cause of cancer deaths. Among various subtypes, the number of patients diagnosed with stage I non-small cell lung cancer (NSCLC), particularly adenocarcinoma, has been increasing. It is estimated that 30 - 40\% of stage I patients will relapse, and 10 - 30\% will die due to recurrence, clearly suggesting the presence of a subgroup that could be benefited by additional therapy. We hypothesize that current attempts to identify stage I NSCLC subgroup failed due to covariate effects, such as the age at diagnosis and differentiation, which may be masking the results. In this context, to stratify stage I NSCLC, we propose CEM-Co, a model-based clustering algorithm that removes/minimizes the effects of undesirable covariates during the clustering process. We applied CEM-Co on a gene expression data set composed of 129 subjects diagnosed with stage I NSCLC and successfully identified a subgroup with a significantly different phenotype (poor prognosis), while standard clustering algorithms failed.
\\
\textbf{Contact:} \href{andrefujita@usp.br}{andrefujita@usp.br}}


\section{Introduction}
\label{sec:introduction}
Lung cancer is the most common malignancy worldwide, with an estimated 228\,150 new cases, and 142\,670 deaths in 2019 only in the United States (American Cancer Society). There are two main types of lung cancer, namely non-small cell lung cancer (NSCLC) and small cell lung cancer (SCLC), which are approximately 80\% and about 10 to 15\% of lung cancers, respectively. 

\begin{figure}
\centering
\includegraphics[width=0.8\textwidth]{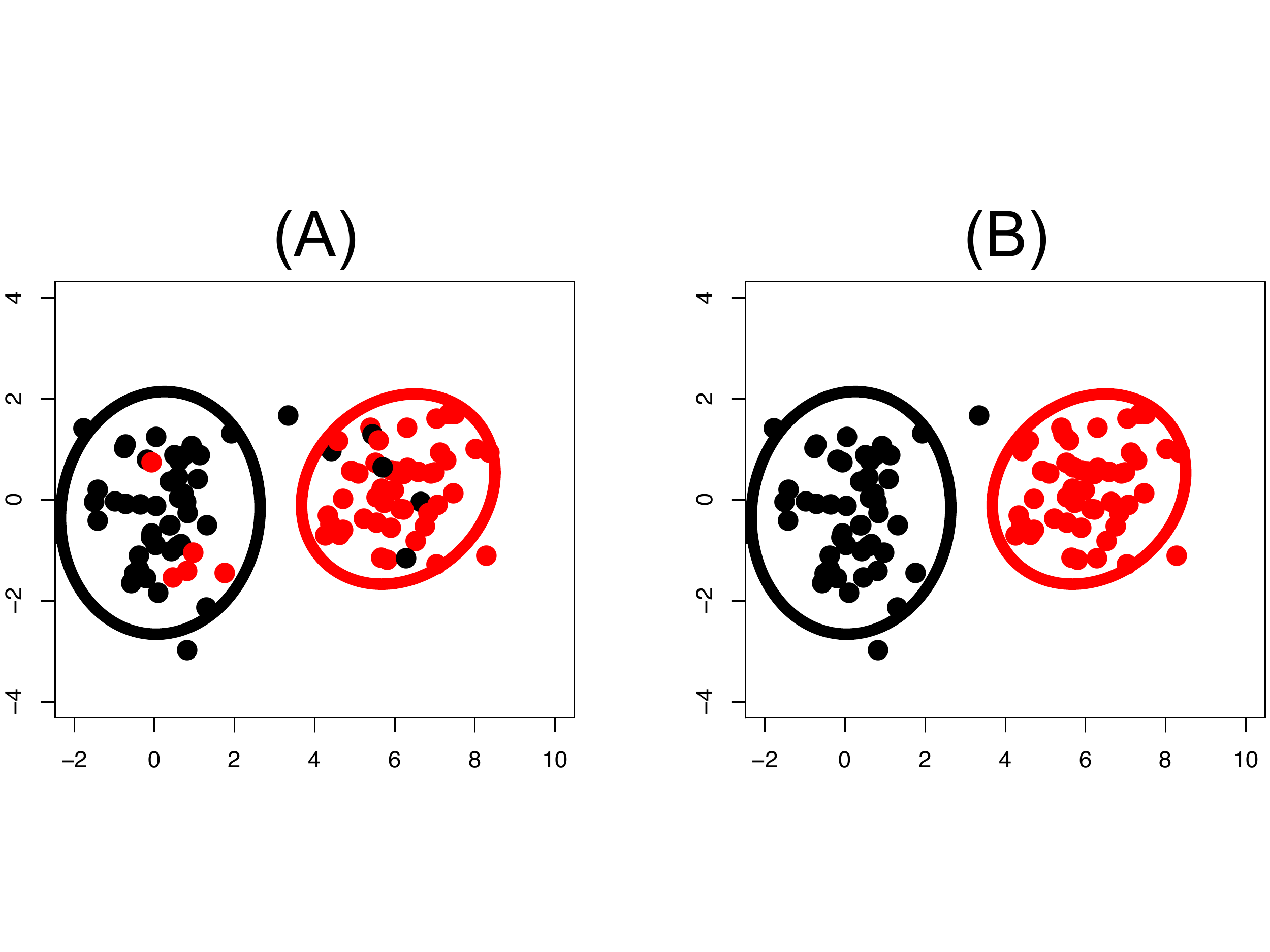} 
\caption{Schema to illustrate the covariate effect on the clustering structure. Suppose two populations of subjects sampled from normal distributions with means $(20, 30)$ and $(30, 35)$, and identity covariance matrices. Moreover, consider that there is a covariate effect on the clustering centroids, such as age, as follows: $(0.1, 0.1)$ on the first cluster and $(-0.2, -0.05)$ on the second cluster. In this case, whether we do not take into account the covariate effect during the clustering process, we cluster some ``red'' items together with the ``black'' items and vice-versa. By considering the covariate effect, we expect to obtain the clustering structure illustrated in panel B.}\label{fig:example}
\end{figure}

Among the patients diagnosed with stage I NSCLC, it is estimated that 30 - 40\% will relapse, and 10 - 30\% will die due to recurrence. This data suggests the presence of at least one subgroup that could be benefited by additional therapy, whether appropriately identified. Thus, further stratification and consequent identification of these individuals become necessary. We hypothesize that one of the reasons that current attempts to stratify stage I NSCLC fail is due to covariate effects, such as the age at diagnosis (and many others), which is widely known to be associated with this disease. To illustrate this hypothesis, see an example in Figure \ref{fig:example}. Suppose there are two groups, i.e., controls and patients. Due to strong covariates effects, such as age at diagnosis, some individuals of the control group (red dots) are assigned to the patients' group (black dots) and vice-versa (Figure \ref{fig:example}A). In this case, the clustering algorithm separated the items into young and elderly. However, what we wished to obtain is one group of controls and another of the patients (Figure \ref{fig:example}B). Therefore, our major problem consists of removing/minimizing the effects of undesirable clinicopathological covariates (e.g., age at diagnosis, gender, adjuvant therapy, and differentiation) that would be masking the results during stratification of stage I NSCLC.

There are several approaches to tackling the problem of clustering by minimizing the covariates' effects. One of the most commons is the one-step approach (a latent class analysis (LCA) based method), which removes the effects of the covariates from the probability of each item to belong to each cluster \cite{bandeen1997latent,dayton1988concomitant}. The problem is that if the strength of the associations between latent class indicators and latent classes is not sufficiently strong, then the parameters of the first-phase LCA model may be affected by auxiliary variables and models \cite{asparouhov2014auxiliary, kamata2018evaluation, nylund2014latent,  vermunt2010latent}. To overcome this problem, the three-step approach uses a mixture model to cluster the data and logistic regression analysis to infer the associations between the predicted class membership and covariates \cite{gudicha2013mixture}. The drawback is that it considers only categorical items, which makes the three-step approach inappropriate for gene expression data analysis (which are continuous). Other approaches basis on the latent profile analysis, which considers the items as continuous data. However, they assume local independence and, therefore, a specific covariance matrix structure \cite{bolck2004estimating},\cite{vermunt2005latent}.

Moreover, they model the covariate effect on the probability of each item belonging to each cluster, but not on the clusters' centroid, where we would expect the covariate effect. Therefore, we are not able to obtain the clustering structure filtered out by the effect of the covariates. Methods to obtain the clustering structure without the effects of covariates usually consider the covariates as additional dimensions of the items or remove the effects of the covariates before the clustering process by carrying out a linear regression. In both cases, the methods assume that the covariates equally affect the clusters' centroids, which may not be true (e.g., age and gender are associated with cancer in different manners).

In this context, we developed a framework namely CEM-Co, which is composed of (i) a model-based clustering algorithm that takes into account the covariates effects (on the centroids and/or covariances), (ii) a statistical test to identify which covariates are associated with the clustering structure, and (iii) a Bayesian information criterion (BIC) to estimate the number of clusters.


\section{Methods}

\label{sec:methods}
Let $N$, $K$, and $P$ be the number of items, clusters, and covariates, respectively. Also, let ${\bf x}_{i}$ be the $i$th $M$-dimensional item ($i=1, \ldots, N$), and ${\bf z}_i=({\bf z}_{1,i}$, ${\bf z}_{2,i}$, $\ldots, {\bf z}_{P,i})$ be the covariates associated with ${\bf x}_{i}$. Then, our goal is to cluster the $N$ items (${\bf x}_1, {\bf x}_2, \ldots, {\bf x}_N$) into $K$ clusters by considering the $P$ covariates (${\bf z}_1, {\bf z}_2, \ldots, {\bf z}_P$).


\subsection{CEM-Co}
\label{sec:cocem}
Similar to the clustering expectation-maximization (CEM) algorithm \cite{celeux1992classification,celeux1995gaussian}, CEM-Co assumes that each cluster is represented by a normal distribution, and therefore the entire data set can be modeled by a mixture of normal distributions.

Let $\mun_{j}^{*}$ ($j = 1, \ldots, K$) be the clusters centroids (means of the normal distributions) without the covariates' effects, $\beta_{j,l}$ be the $M$-dimensional coefficient representing the strength of the $l$th covariate effect ($l=1,\ldots,P$) on the $j$th cluster centroid. Then, we model the ``observed'' clusters centroids as:


\begin{equation}
\label{eq:centroid}
\mun_{i,j} | {\bf z}_{i} = \mun_{j}^{*} + \betan_{j,1} {\bf z}_{1,i} + \betan_{j,2} {\bf z}_{2,i} + \ldots + \betan_{j,P} {\bf z}_{P,i}.
\end{equation}

Let $\mathbf{L}_{i,j}$ be a diagonal $M \times M$ matrix whose the $r$th diagonal ($r=1, \ldots, M$) is $\sigma_{r,j} + \gamma_{1,r,j} {\bf z}_{1,i}+\gamma_{2,r,j} {\bf z}_{2,i}+\ldots+\gamma_{P,r,j} {\bf z}_{P,i}$, and $\mathbf{E}_{j}$ be a $M \times M$ positive definite matrix. Then, we model the covariance matrix of the multivariate normal distribution (cluster) as:

\begin{equation}
\label{eq:covariance}
\Sigman_{i,j} = \mathbf{L}_{i,j} \mathbf{E}_{j} \mathbf{L}_{i,j}.
\end{equation}

Notice that $\Sigman_{i,j}$ is in the form of a Choleski decomposition \cite{dereniowski2003cholesky}, which implies in a positive-definite matrix. If there is no covariate effect on the covariance matrix $\Sigman_{i,j}$, then $\sigma_{r,j}=1$ and $\gamma_{l,r,j}=0$ for all $l=1, \ldots, P$, $r=1, \ldots, M$, and $j=1, \ldots, K$. In other words, $\mathbf{L}_{i,j}$ is an identity matrix and, consequently, the covariance matrix is defined only by $\mathbf{E}_{j}$.

Let 

\begin{equation}
\label{eq:phi}
\phi({\bf x}_i, \mu_{i,j}, \Sigman_{i,j})=\frac{\exp(-\frac{1}{2}({\bf x}_i-\mu_{i,j})^{\top}\Sigman_{i,j}^{-1}({\bf x}_i-\mu_{i,j}))}{\sqrt{\textrm{det}(2\pi\Sigman_{i,j})}},
\end{equation}

be the density of a multivariate normal distribution, and $\alpha_{j}$ be the weight associated with the $j$th cluster ($\sum_{j=1}^{K} \alpha_{j} = 1$). Then we can write the likelihood function as:


\begin{equation}
\label{eq:likelihood}
L(\mathbf{\theta} | {\bf x}_1, \ldots, {\bf x}_N) = \prod_{i=1}^{N} \sum_{j=1}^{K} \alpha_{j} \phi({\bf x}_i, \mu_{i,j}, \Sigman_{i,j}).
\end{equation}

where $\mathbf{\theta} = (\mathbf{\alpha}, \mathbf{\mu^*}, \mathbf{\beta}, \mathbf{L}, \mathbf{E})$.
Based on Equations \ref{eq:centroid}, \ref{eq:covariance}, \ref{eq:phi}, and \ref{eq:likelihood}, we can simultaneously estimate the parameters ($\mu_{j}^{*}$, $\mathbf{E}_{j}$, $\sigma_{r,j}$, $\gamma_{l,r,j}$, $\betan_{j,l}$, $\alpha_{j}$) and carry out the clustering by using an expectation-maximization (EM) algorithm described in Algorithm \ref{alg:cocem}.

\RestyleAlgo{boxruled}
\LinesNumbered
\begin{algorithm}[ht]
  \caption{Clustering expectation-maximization with covariates effects on both the clusters centroids and covariance matrices  \newline 
  		  {\bf Input}: the items ${\bf x}_i$ ($i=1, \ldots, N$), the number of clusters $K$, and the covariates ${\bf z}_{l,i}$ ($l=1, \ldots, P$).\newline
          {\bf Output}: the $K$ clusters \label{alg:cem}}\label{alg:cocem}

	Let $\mu_{i,j}$ (Equation \ref{eq:centroid}), $\Sigman_{i,j}$ (Equation \ref{eq:covariance}), $\alpha_j$, and $\phi({\bf x}_i, \mu_{i,j}, \Sigman_{i,j})$ (Equation \ref{eq:phi}) be the centroid, the covariance matrix, the weight associated with the $j$th normal distribution ($j=1, \ldots, K$), and the density of the multivariate normal distribution, respectively.
    
    Randomly initialize the parameters of the $K$ normal distributions, i.e., the centroids ($\mu_{j}^{*}$), covariances matrices ($\mathbf{E}_{j}$, $\sigma_{r,j}$, $\gamma_{l,r,j}$), covariates effects ($\betan_{j,l}$), and weights for each normal distribution ($\alpha_{j}$) $(r=1, \ldots, M)$.\label{step:init}
    
	{\it Expectation step}. Compute the expected probability for the $i$th item to belong to the $j$th cluster as $P_{i,j} = \frac{\hat{\alpha}_j \phi({\bf x}_i, \hat{\mu}_{i,j}, \hat{\Sigman}_{i,j})}{\sum_{l=1}^{k}\hat{\alpha}_l \phi({\bf x}_i, \hat{\mu}_{i,l}, \hat{\Sigman}_{i,l})}$. \label{E-step}
    
    {\it Clustering step}. Let ${\bf C}$ be a $(N \times K)$ matrix where ${\bf C}_{i,j}=1$ if ${\bf x}_i$ belongs to the $j$th cluster, and ${\bf C}_{i,j}=0$, otherwise. Update ${\bf C}$ by assigning each $i$th item to the $j$th cluster, which provides the maximum current expected probability $P_{i,j}$.

    {\it Maximization step}. Let $\hat{d}_{i,j} = \hat{\betan}_{j,1} {\bf z}_{1,i} + \hat{\betan}_{j,2} {\bf z}_{2,i} + \ldots + \hat{\betan}_{j,P} {\bf z}_{P,i}$ be the sum of all covariates effects. Then, compute the maximum likelihood estimates for ${\hat \mun}_{j}^{*} = \frac{\sum_{i=1}^{N} (\xn_{i}-\hat{d}_{i,j}) \hat{P}_{i,j}} {\sum_{i=1}^{N} \hat{P}_{i,j}}$, $\hat{\mathbf{E}}_{j} = \frac{\sum_{i=1}^{N}\hat{P}_{i,j} (\hat{\mathbf{L}}_{i,j}^{-1} (\xn_{i}-\hat{\mun}_{j}^{*}-\hat{d}_{i,j})) (\hat{\mathbf{L}}_{i,j}^{-1}(\xn_{i}-\hat{\mun}_{j}^{*}-\hat{d}_{i,j}))^{\top}}{\sum_{i=1}^{N} \hat{P}_{i,j}}$, $\hat{\betan}_{j,l} = \frac{\sum_{i=1}^{N} {\bf z}_{l,i} (\xn_{i}-\hat{\mun}_{j}^{*}-\hat{d}_{i,j}+\hat{\betan}_{j,l} {\bf z}_{l,i}) \hat{P}_{i,j}} {\sum_{i=1}^{N} {\bf z}_{l,i}^{2} \hat{P}_{i,j}}$, and ${\hat \alpha}_{j} = \frac{\sum_{i=1}^{N}{\bf C}_{i,j}}{N}$. To estimate the maximized value of the likelihood function, find the roots (or zeros) of the partial derivative of $\hat{\mathbf{L}}_{i,j}$ by using the Newton-Raphson method.
    
	Go to step \ref{E-step} until convergence of the likelihood function (Equation \ref{eq:likelihood}).
\end{algorithm}

Similar to the CEM algorithm, CEM-Co depends on the initialization of the parameters in step \ref{step:init} of Algorithm \ref{alg:cocem} ($\alpha_{j}, \mu^*_{j}, \beta_{j,l}, \mathbf{L}_{i,j}, \mathbf{E}_{j}$, for $i=1, \ldots, N$, $j=1, \ldots, K$, and $l=1, \ldots, P$). Different initializations may lead to different likelihood values (EM algorithm usually obtains a local optimum). Thus, we suggest to run Algorithm \ref{alg:cocem} several times with different initialization values and select the one with the highest likelihood.

\subsection{Statistical test for the covariate effect}
\label{sec:LRT}

In some empirical data analyses, one may be interested in testing whether a covariate is statistically associated with the clustering structure. 

For example, suppose we want to test whether the $l$th covariate (${\bf z}_{l}$) has no effect on the clustering structure. In other words, consider the following statistical test:

$H_0: \betan_{1,l} = \betan_{2,l} = \ldots = \betan_{K,l} = \mathbf{0}$

versus

$H_1$: At least one $\betan_{j,l} \neq \mathbf{0}$ ($j=1, \ldots, K$).

To this end, we propose a likelihood ratio test (LRT) \cite{wilks1938large}.

Let $L(\alpha_{j}, \mu^*_{j}, \beta_{j,l}, \mathbf{L}_{i,j}, \mathbf{E}_{j}| {\bf x}_1, \ldots, {\bf x}_N, \betan_{1,l} = \ldots = \betan_{K,l} = \mathbf{0})$ be the likelihood function assuming that the $l$th covariate has no effect on the $j$th centroid (the null model), and $L(\alpha_{j}, \mu^*_{j}, \beta_{j,l}, \mathbf{L}_{i,j}, \mathbf{E}_{j} | {\bf x}_1, \ldots, {\bf x}_N))$ be the likelihood function for the alternative model (Equation \ref{eq:likelihood}). Then, we can define the statistic of the test ($D$) as:

\begin{equation}
D = 2\{\ln(L(\mathbf{\hat{\theta}} | {\bf x}_1, \ldots, {\bf x}_N)) - \\ \ln(L(\mathbf{\hat{\theta}_{0}}| {\bf x}_1, \ldots, {\bf x}_N)) \}
\end{equation}
where $\mathbf{\hat{\theta}_{0}}$ are the coefficients obtained by the maximum likelihood estimator under $H_{0}$ ($\betan_{1,l} = \betan_{2,l} = \ldots = \betan_{K,l} = \mathbf{0}$).

To obtain $D$, carry out the CEM-Co algorithm with (the alternative model) and without (the null model) the $l$th covariate and compare the two likelihood functions.

The probability distribution of the test statistic $D$ is approximately a chi-squared distribution with degrees of freedom equal to the number of parameters of the alternative model minus the number of parameters of the null model. For example, in our specific test ($H_0: \betan_{1,l} = \betan_{2,l} = \ldots = \betan_{K,l} = \mathbf{0}$), we have $K$ parameters to be estimated for all $\alpha_{j}$, $KM$ for all $\mu^*_{j}$, $KMP$ for all $\betan_{j,l}$, $KM^2$ for all $\mathbf{E}_{j}$, and $2KM$ for all $\mathbf{L}_{i,j}$. Thus, the total number of parameters is $K(1 + M(3+P) + M^{2})$. To test the effect of a single covariate, we have $K(1 + M(2+P) + M^{2})$ parameters under $H_{0}$. Therefore, the number of degrees of freedom for this test is $KM$.

We may use a similar procedure to test the effect of the $l$th covariate (${\bf z}_{l,i}$) on the covariance matrix, i.e.,

$H_0: \gamma_{l,1,1} = \gamma_{l,1,2} = \ldots = \gamma_{l,1,K} = \gamma_{l,2,1} = \ldots = \gamma_{l,M,K} =  \mathbf{0}$

versus

$H_1$: At least one $\gamma_{l,r,j} \neq  \mathbf{0}$ ($j=1, \ldots, K$ and $r=1, \ldots, M$).

Again, to obtain $D$, carry out the CEM-Co algorithm with (the alternative model) and without (the null model) the $l$th covariate and compare the two likelihood functions. The number of degrees of freedom to this case is also $KM$ (there are $KM$ parameters $\gamma_{l,1,1}, \ldots, \gamma_{l,1,K}, \gamma_{l,2,1}, \ldots, \gamma_{l,M,K}$).

To use the chi-squared distribution in the LRT, the likelihood function must follow the regularity conditions \cite{hogg2005introduction}. The proof for these conditions is straightforward for the tests described in the previous paragraphs, including the fact that, under the null hypothesis, the parameters $\betan_{i,l}$ and $\gamma_{l,M,K}$ cannot be in the border of the distribution support. Notice that, under $H_0$, this condition is valid because $\betan_{i,l}$ and $\gamma_{l,M,K}$ are real numbers for any $i, l, M$, and $K$.



\subsection{Estimation of the number of clusters}
\label{sec:BIC}

As described in Algorithm \ref{alg:cocem}, CEM-Co requires the number of clusters ($K$) as input. However, in empirical data analysis, we rarely know the number of clusters {\it a priori}; thus, we must estimate it. To estimate the number of clusters, we propose to use the Bayesian information criterion (BIC). Let $\hat{L}_K$ be the maximized value of the likelihood function (Equation \ref{eq:likelihood}) of the clustering structure obtained by CEM-Co with $K$ clusters, $N$ be the number of items, and $R=K(1 + (3+P)M + M^{2})$ be the number of parameters estimated by the model. Then, we define the BIC for the clustering structure obtained by CEM-Co with $K$ clusters as:

\begin{equation}
\label{eq:BIC}
{\displaystyle \mathrm {BIC}_K ={\ln(N)R-2\ln({\hat {L}_K})}.\ }
\end{equation}

The estimated number of clusters $\hat{K}$ is the one, which minimizes the BIC statistic (Equation \ref{eq:BIC}).


\subsection{CEM-Co with nonlinear effect}
Sometimes the covariates are nonlinearly associated with the clusters' centroids (e.g., quadratic or sigmoid relationships). In this case, the method presented in section \ref{sec:cocem} may not be suitable. To model nonlinear relationships, we extended CEM-Co as follows. 

Let $f_{j,l}$ be a function representing the strength of the $l$th covariate effect on the $j$th cluster centroid. Then we can write the ``observed'' clusters centroids as :

\begin{equation}
\mun_{i,j} = \mun_{j}^{*} + f_{j,1} ({\bf z}_{1,i}) + f_{j,2} ({\bf z}_{2,i}) + \ldots + f_{j,P} ({\bf z}_{P,i}).
\end{equation}

To model $f_{j,l}$, we use B-spline \cite{de1972calculating}. Let $S$ be the number of knots and $B$ be the polynomial degree of the B-spline, then we represent $f_{j,l}$ by a matrix $\mathbf{W}_{j,l}$ (with dimensions ($N \times (S+B)))$, where the $i$th row represents the spline basis of the $l$th covariate effect on the $i$th item in the $j$th cluster. We define the matrix $\mathbf{W}_{j}$ $(N \times (P(S+B)))$ as the combination of $\mathbf{W}_{j,1}, \mathbf{W}_{j,2}, \ldots, \mathbf{W}_{j,P}$ by columns. Notice that we model the nonlinear effect only for the centroids, and not for the covariance matrices.

To obtain the CEM-Co algorithm with nonlinear covariates effects, replace ${\bf z}_{l}$ by the B-spline basis represented by $\mathbf{W}_{j,l}$ in Algorithm \ref{alg:cocem}. To statistically test the effect of a single covariate, consider the number of degrees of freedom as $KM(S+B)$ in the likelihood ratio test (section \ref{sec:LRT}). To estimate the number of clusters (section \ref{sec:BIC}), consider the number of parameters as $K(1 + M(3+P(S+B)) + M^{2})$. 


\section{Simulations}







\label{sec:simulation}
To evaluate the performance of CEM-Co in clustering the items, the power of the LRT, and the accuracy of the BIC in estimating the number of clusters, we designed three scenarios: (1) the covariates are linearly associated with the clusters' centroids, (2) the covariates are associated with the covariance matrix, and (3) the covariates are nonlinearly associated with the clusters' centroids.

{\it Scenario 1}. Covariates with linear effects on the clusters' centroids.

\begin{enumerate}
\item Set the number of clusters ($K=2$), the number of dimensions ($M=5$), and the number of covariates ($P=5$).\label{step:1}

\item Set the centroids (means) of the $K=2$ clusters to $\mu_1^{*}=(0,0,0,0,0)$ and $\mu_2^{*}=(0.2,0.2,0.2,0.2,0.2)$.\label{step:2}

\item Set the effects of the covariates on each cluster ($\beta$) as the same as estimated by CEM-Co in the real data set (Section \ref{sec:lung}).\label{step:3}

\item For each item ($i=1, \ldots, N$), simulate $P=5$ covariates, two from a normal distribution with zero mean and unit variance (${\bf z}_{1,i} \sim N(0,1)$, ${\bf z}_{2,i} \sim N(0,1)$) and three from binomial distributions with parameters equal to $0.4, 0.25$, and $0.15$ (${\bf z}_{3,i} \sim B(0.4)$, ${\bf z}_{4,i} \sim B(0.25)$, ${\bf z}_{5,i} \sim B(0.15)$). \label{step:4}

\item Simulate $N/2$ items for each cluster from a multivariate normal distribution with means equal to $\mu_{i,j} = \mu_{j}^{*} + \beta_{1,j} {\bf z}_{1,i} + \beta_{2,j} {\bf z}_{2,i} + \beta_{3,j} {\bf z}_{3,i} + \beta_{4,j} {\bf z}_{4,i} + \beta_{5,j} {\bf z}_{5,i}$ (for $i=1, \dots, N/2$ and $j=1,2$) and a covariance matrix equals to a $(5 \times 5)$ identity matrix multiplied by $0.03$ for each cluster.\label{step:5}
\end{enumerate}

{\it Scenario 2}. Covariates with linear effects on the covariance matrices.

\begin{enumerate}
\item Set the number of clusters ($K=4$), the number of dimensions ($M=2$), and the number of covariates ($P=1$).\label{step:1,scen:2}

\item Set the centroids (means) of the $K=4$ clusters to $\mu_1^{*}=(0,0)$, $\mu_2^{*}=(0,1)$, $\mu_3^{*}=(1,0)$, and $\mu_4^{*}=(1,1)$.\label{step:2,scen:2}

\item Set the effects of the covariate on each cluster to $\beta_{1}=(0.3,0.3)$, $\beta_{2}=(-0.3,-0.3)$, $\beta_{3}=(0.3,-0.3)$ and $\beta_{4}=(-0.3,0.3)$.\label{step:3,scen:2}

\item Simulate one covariate for each item ($i=1, \ldots, N$) from a normal distribution with unit mean and unit variance (${\bf z}_{1,i} \sim N(1,1)$).

\item Let $\mathbf{E}_{j}$ be a $(2 \times 2)$ matrix with $0.1$ at the diagonal, $\sigma_{r,j}=1$ (for $r = 1, 2$ and $j = 1, 2, 3, 4$), $\gamma_{1, r, j} = w_{j}$ (for $r = 1, 2$, $w_{1}=1$, $w_{2}=1$, $w_{3}=1$, and $w_{4}=10$), and $\mathbf{L}_{i,j}$ be a $(2 \times 2)$ matrix as defined in section \ref{sec:cocem}. Then, simulate $N/K$ items from a multivariate normal distribution with mean and covariance matrix equal to $\mu_{i,j} = \mu_{j}^{*} + \beta_j {\bf z}_{1,i}$ and $\Sigman_{i,j} = \mathbf{L}_{i,j} \mathbf{E}_{j} \mathbf{L}_{i,j}$, respectively, for each cluster $j=1, \ldots, K$.
\end{enumerate}

{\it Scenario 3}. Covariates with nonlinear effects on the clustering structures. To simulate non-linear covariate effects on the clusters centroids, replace step \ref{step:5} of scenario 1 by:

\begin{enumerate}
\item Let $f_{j,1}({\bf z}_{1,i}) = \beta_j{\bf z}_{1,i}+\beta_j{\bf z}_{1,i}^{2}$ be a quadratic function representing the strength of the covariate effect on the $j$th cluster centroid. Then, simulate $N/K$ items from a multivariate normal distribution with mean equals to $\mu_{i,j} = \mu_{j}^{*} + f_{j,1}({\bf z}_{1,i})$, and a covariance matrix as a $(2 \times 2)$ identity matrix multiplied by $0.1$, for each cluster $j=1, \ldots, K$.
\end{enumerate}

For scenario 1, the number of items varied in $N=120$, $240$, and $360$. Notice that we designed this scenario to be similar to the real data described in section \ref{sec:lung}. For scenarios 2 and 3, the number of items varied in $N=200, 300, 400$, and $800$. For scenario 3, we modeled $f_{j,1}$ with a cubic B-spline with four degrees of freedom. We carried out $300$ repetitions for each scenario and each number of items ($N$).

\section{Stage I lung adenocarcinoma}
\label{sec:lung}

\subsection{Stage I NSCLC data set}
We collected a set of clinically annotated gene expression data composed of 45 genes and 129 subjects diagnosed with stage I non-small cell lung cancer (NSCLC) by real-time RT-PCR. We normalized the gene expression levels relative to the internal housekeeping control 18S gene by using the method described in \cite{schmittgen2008analyzing}. Clinical and pathological characteristics are summarized in Table \ref{tab:clinico}.

 
\begin{table}
\centering
\caption{Clinicopathological characteristics of patients and their tumors. The data are numbers (\%) unless otherwise stated. SD: standard deviation.}\label{tab:clinico}
\begin{tabular}{ccc}
\hline
\hline
Variable&Status&$n$\\
\hline
&Mean (SD)&66.26 (10.04)\\
Age at diagnosis (years)&$\leq65$&60 (46.51)\\
&$\> 65$&69 (53.49)\\\\
Gender&Male&59 (45.74)\\
&Female&74 (54.26)\\\\
&Yes&18 (13.95)\\
Therapy&No&110 (85.27)\\
&Unknown&1 (0.77)\\\\
&1&40 (31.00)\\
Differentiation&2&51 (39.53)\\
&3&38 (29.46)\\\\
Dead&Yes&48 (37.21)\\
&No&81 (62.79)\\
\hline

\end{tabular}
\end{table}

\subsection{Gene expression pre-processing}
We re-scaled gene expression data for zero mean and unit variance. Then, we applied a principal component analysis (PCA) for dimension reduction and selected (by the elbow method) five principal components (PCs) associated with the five largest eigenvalues, which altogether represent 51.87\% of the variance. These five PCs are the representative new dimensions of the subjects diagnosed with stage I NSCLC. Thus, we have $N=129$ individuals, $M=5$ dimensions, and $P=4$ covariates (age at diagnosis, gender, therapy, and differentiation).

\section{Results and Discussions}
To evaluate the performance of CEM-Co, we compared it against three other usual approaches, namely:
\begin{enumerate}
\item CEM: the application of the standard clustering expectation-maximization (CEM) algorithm on the original data set, i.e., without taking into account the covariates;
\item CEM-dimension: the application of the standard CEM algorithm, but considering the covariates as additional dimensions of the items; and
\item CEM-partial: the application of the standard CEM algorithm on the residues of the linear regression model, where the response variable is the item, and the exploratory variables are the covariates.
\end{enumerate}

We did not compare CEM-Co against the three-step approach because the clustering structure is the same obtained by CEM (the three-step approach clusters the data by using CEM and then identifies the association between the items and covariates).

\begin{figure*}
\centering
\includegraphics[width=0.8\textwidth]{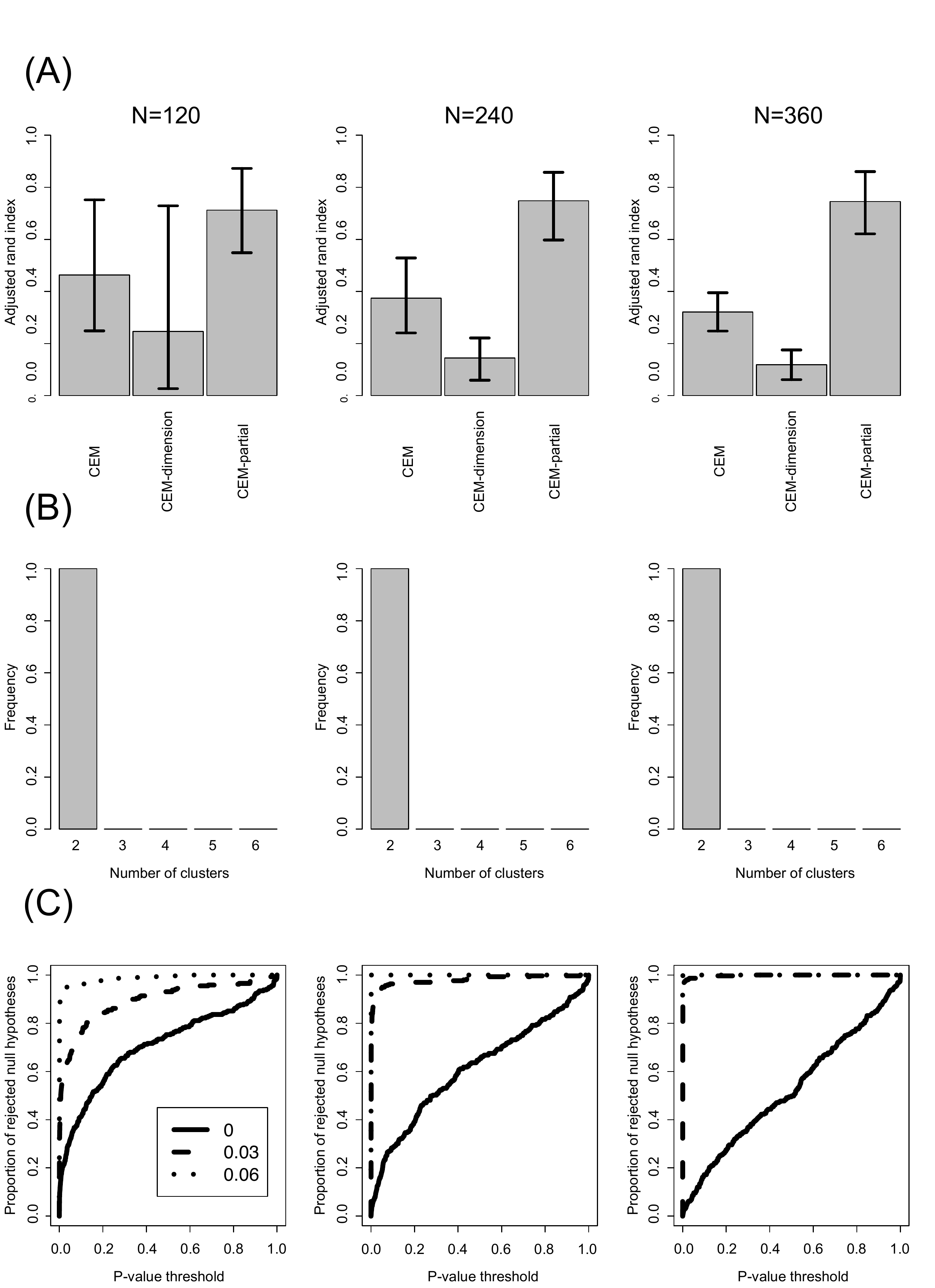} 
\caption{Linear covariate effect on the centroids. (A) Differences between the adjusted rand indices obtained by CEM-Co and the alternative methods (CEM, CEM-dimension, CEM-partial, and one-step). The greater is this difference; the better is the performance of CEM-Co than the alternative method. Error bars represent the 90\% confidence intervals. (B) Estimation of the number of clusters. The bars represent the frequency BIC selected the indicated number of clusters. For all evaluated number of items ($N$), BIC selected the number of clusters correctly as two. (C) ROC curves. The area below the curve represents the power of the statistical test. The solid line represents the covariate effect under the null hypothesis ($\beta=0$). Dashed lines represent the test under the alternative hypothesis ($\beta > 0$). The power of the LRT increases proportionally to the number of items ($N$) and covariate effect strength. For a small number of items ($N=120, 240$), the LRT did not control for the type I error (the solid line is above the diagonal).}\label{fig:cocem_linear}
\end{figure*}

First, we analyzed the case the covariates are linearly associated with the centroids. To evaluate how well the algorithms cluster the items, we calculated the adjusted rand index for each repetition. Since we carried out all the four algorithms on the same set of items, we present the differences between the adjusted rand indices obtained by CEM-Co minus the one obtained by each of the three alternative methods (Figure \ref{fig:cocem_linear}A). The greater is this difference; the better is CEM-Co than the alternative approach. Figure \ref{fig:cocem_linear}A shows that the performance of CEM-Co is better than all the alternative approaches. Figure \ref{fig:cocem_linear}B shows the frequency the BIC selected each number of clusters. As expected, the BIC accurately estimated the correct number of clusters ($K=2$) for all examined sample sizes. Figure \ref{fig:cocem_linear}C shows the ROC curves obtained by testing the covariate effect. To verify the control of the type I error, we simulated a data set as described in Section \ref{sec:simulation} - scenario I, but under the null hypothesis, i.e., when there is no covariate effect. Observe that when $\beta=0$ (no covariate effect), the ROC curve (solid line) should be at the diagonal. Indeed, for $N=360$, the LRT effectively controlled the type I error (the solid line is at the diagonal). However, for $N=120, 240$, the ROC curves (solid lines) are above the diagonal, i.e., it rejected the null hypothesis more than the expected by the p-value threshold. Then, for small $N$, instead of using the LRT, we propose to use a parametric bootstrap procedure. For $N=120$ and p-value thresholds at $0.01, 0.05$, and $0.10$, the empirical false-positive rates for the bootstrap-based test (with $1\,000$ bootstrap samples) were $0.01, 0.07$, and $0.16$, respectively, while for the LRT were $0.06, 0.17$, and $0.27$. Thus, for small $N$, the type I error is better controlled by the bootstrap procedure than the LRT. Under the alternative hypothesis, the proportion of rejected null hypothesis (power of the test) by the LRT increases as the number of items ($N$) and/or the covariate effect increases (dashed lines). 

\begin{figure*}
\centering
\includegraphics[width=0.8\textwidth]{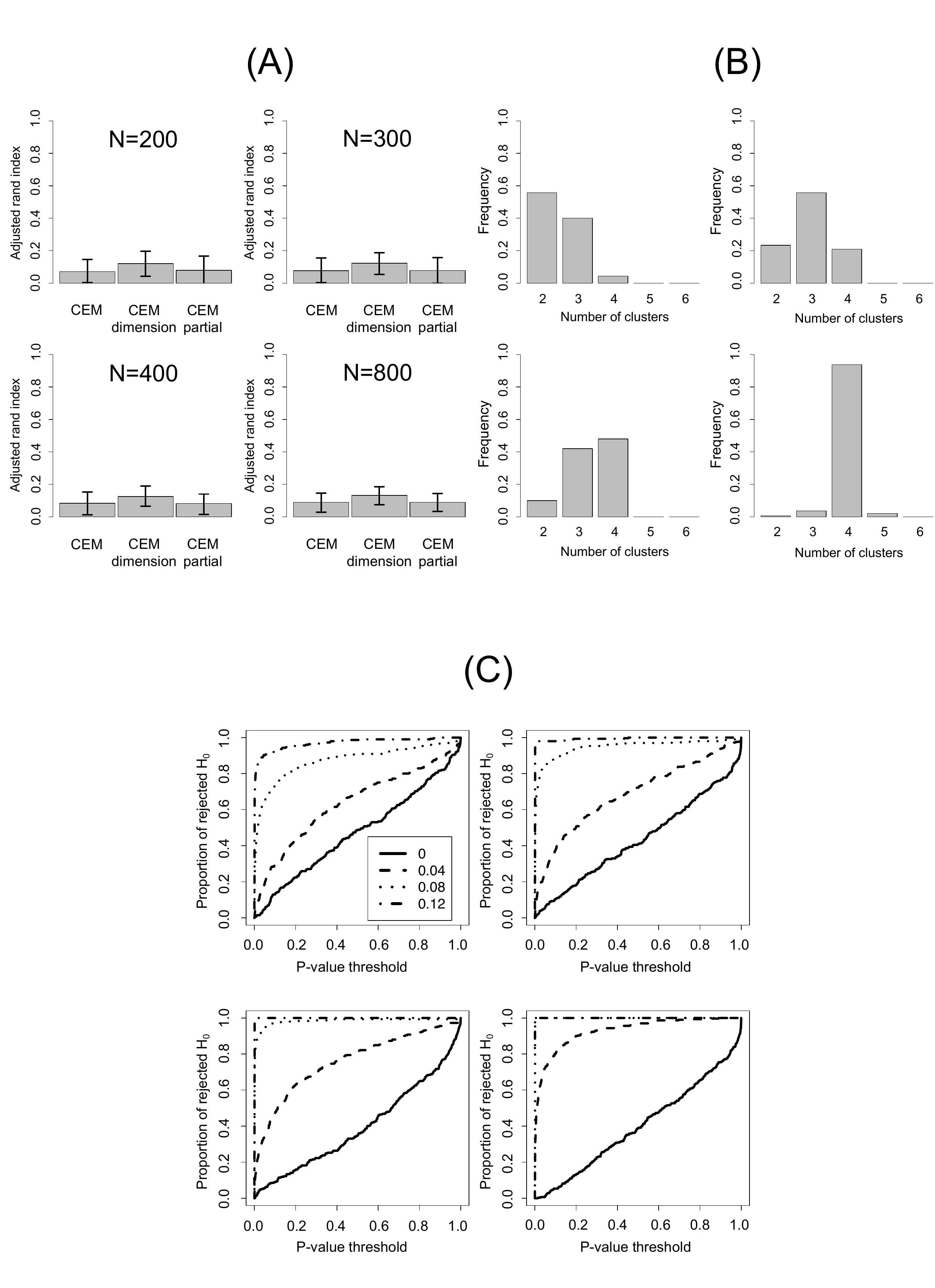} 
\caption{Linear covariate effect on the covariances. (A) Differences between the adjusted rand indices obtained by CEM-Co and the alternative methods (CEM, CEM-dimension, CEM-partial, and one-step). The greater is this difference; the better is the performance of CEM-Co than the alternative methods. Error bars represent the 90\% confidence intervals. (B) Estimation of the number of clusters. The bars represent the frequency BIC selected the indicated number of clusters. As the number of items increases ($N$), BIC converges to the correct number of clusters ($K=4$). (C) ROC-like curves. The area below the curve represents the power of the statistical test. The solid line represents the covariate effect under the null hypothesis ($\beta=0$). Dashed lines represent the test under the alternative hypothesis ($\beta > 0$). The power of the likelihood ratio test increases proportionally to the number of items ($N$) and covariate effect strength.}\label{fig:cocem_variance}
\end{figure*}

We also analyzed the case the covariate affects the covariance. Figure \ref{fig:cocem_variance}A shows that the performance of CEM-Co is better than all the alternative approaches. Figure \ref{fig:cocem_variance}B shows that the BIC accurately estimated the correct number of clusters ($K=4$) as the sample size increases. Figure \ref{fig:cocem_variance}C shows that, under the null hypothesis, the LRT effectively controlled the type I error (the solid line is at the diagonal). Under the alternative hypothesis, the proportion of rejected null hypothesis (power of the test) by the LRT increases as the number of items ($N$) and/or the covariate effect increases (dashed lines).

\begin{figure*}
\centering
\includegraphics[width=0.8\textwidth]{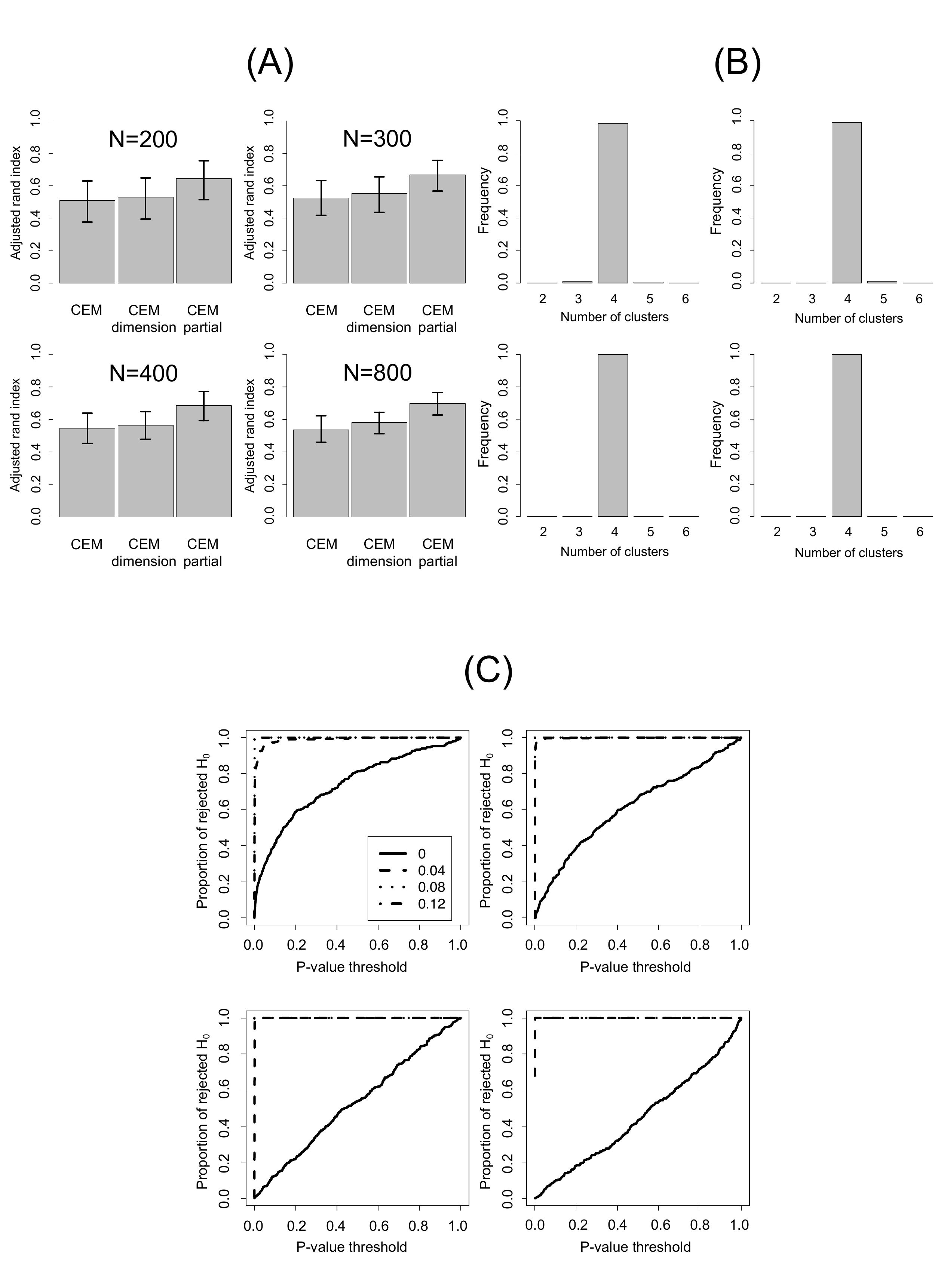} 
\caption{Nonlinear covariate effect on the centroids. (A) Differences between the adjusted rand indices obtained by CEM-Co and the alternative methods (CEM, CEM-dimension, and CEM-partial). The greater is this difference; the better is the performance of CEM-Co than the alternative methods. Error bars represent the 90\% confidence intervals. (B) Estimation of the number of clusters. The bars represent the frequency BIC selected the indicated number of clusters. For all evaluated number of items ($N$), BIC selected the number of clusters correctly as four. (C) ROC-like curves. The area below the curve represents the power of the statistical test. The solid line represents the covariate effect under the null hypothesis ($\beta=0$). Dashed lines represent the test under the alternative hypothesis ($\beta > 0$). The power of the likelihood ratio test increases proportionally to the number of items ($N$) or covariate effect strength.}\label{fig:cocem_nonlinear}
\end{figure*}

Besides, we analyzed the case the covariates are nonlinearly (e.g., quadratically) associated with the items. Figure \ref{fig:cocem_nonlinear}A shows that the performance of CEM-Co is better than all the alternative approaches. Figure \ref{fig:cocem_nonlinear}B shows that the BIC accurately estimated the correct number of clusters ($K=4$) for all evaluated sample sizes. Figure \ref{fig:cocem_nonlinear}C shows that, under the null hypothesis, the LRT effectively controlled the type I error (the solid line is at the diagonal) as the sample size increases. Under the alternative hypothesis, the proportion of rejected null hypothesis (power of the test) by the LRT increases as the number of items ($N$) and/or the covariate effect increases (dashed lines).

Finally, we applied CEM-Co and the other three alternative methods on the stage I NSCLC gene expression data. We only analyzed the case that the effect of covariates may be linearly associated with the centroids. We did not analyze the effect of covariates on the covariances nor the nonlinear case because the number of parameters becomes higher than the number of items/individuals. We considered the age at diagnosis, gender, adjuvant therapy, and differentiation as covariates.

\begin{figure}
\centering
\includegraphics[width=0.6\textwidth]{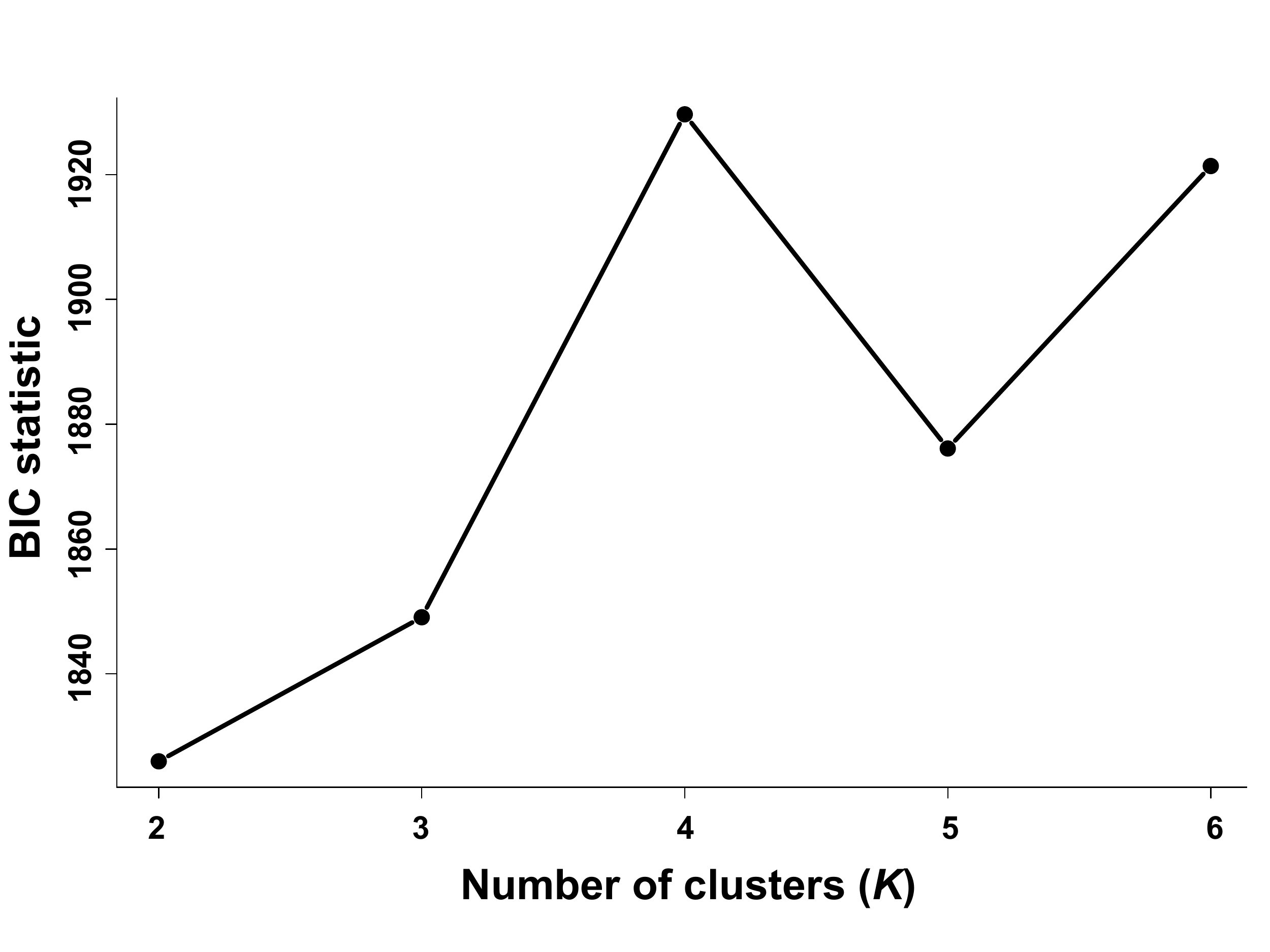} 
\caption{Estimation of the number of clusters in stage I NSCLC gene expression data by using the BIC for CEM-Co. The BIC statistic reaches its minimum value for $K=2$. Therefore, the estimated number of clusters in this data set is ${\hat K}=2$.}\label{fig:BIC}
\end{figure}

To estimate the number of clusters, we used the BIC described in section \ref{sec:BIC}. The estimated number of clusters was two for all clustering algorithms (Figure \ref{fig:BIC}).

\begin{figure}
\centering
\includegraphics[width=0.6\textwidth]{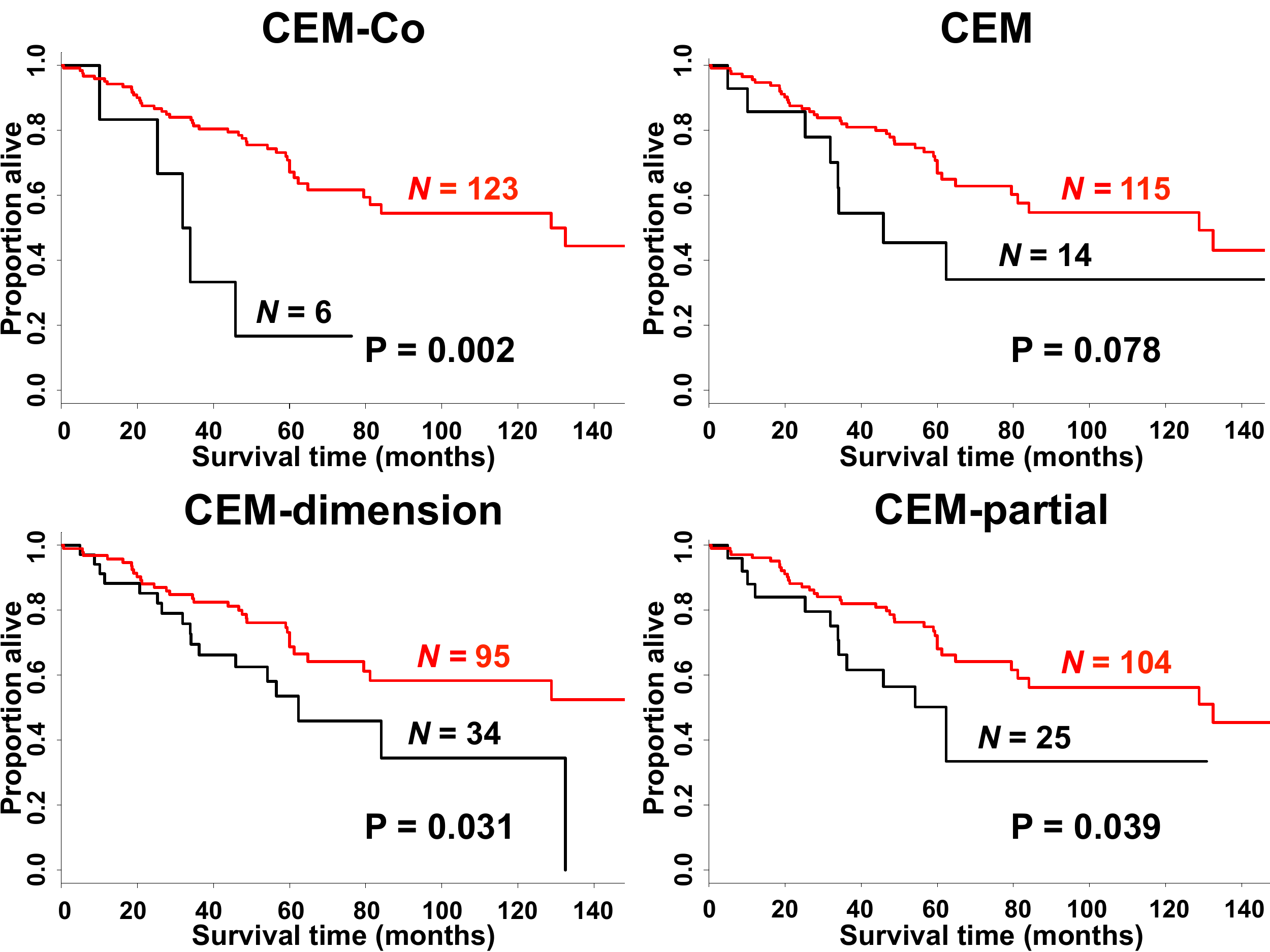} 
\caption{Kaplan-Meier survival curves constructed based on the clusters obtained by the application of CEM-Co, CEM, CEM-dimension, and CEM-partial, on the stage I NSCLC gene expression data set. The p-values were obtained by the log-rank test.}\label{fig:KM}
\end{figure}

After clustering the individuals by using CEM-Co, CEM, CEM-dimension, and CEM-partial, we verified whether the obtained two clusters present different phenotypes. One clinically relevant outcome in cancer is survival time. Then, we analyzed the association of these two clusters with the survival outcome using a Kaplan-Meier non-parametric curve (Figure \ref{fig:KM}) and a multivariate Cox proportional hazards model with age at diagnosis, gender, adjuvant therapy, and differentiation as covariates (Table \ref{tab:cox}).

\begin{table}
\centering
\caption{Multivariate survival analyses were performed based on the Cox regression model. CI: confidence interval. P-values $<$ 0.05 are in bold.}\label{tab:cox}
\begin{tabular}{cccc}
\hline
\hline
Method&Variables&Hazard ratio (95\%CI)&P-value\\
\hline
&Cluster (Red/Black)&0.117 (0.041 - 0.334)&{\bf $<$0.001}\\
&Age at diagnosis&1.084 (1.048 - 1.121)&{\bf $<$0.001}\\
CEM-Co&Gender (male/female)&1.231 (0.668 - 2.268)&0.5045\\
&Therapy (yes/no)&2.545 (1.223 - 5.296)&{\bf 0.013}\\
&Differentiation (2/1)&1.348  (0.605 - 3.003)&0.466\\
&Differentiation (3/1)&2.015 (0.853 - 4.762)&0.1101\\

\hline
&Cluster (Red/Black)&0.436 (0.181 - 1.052)&0.0646\\
&Age at diagnosis&1.082 (1.045 - 1.120)&{\bf $<$0.001}\\
CEM&Gender (male/female)&1.220 (0.667 - 2.232)&0.519\\
&Therapy (yes/no)&2.151 (1.049 - 4.410)&{\bf 0.037}\\
&Differentiation (2/1)&2.064 (0.701 - 3.438)&0.278\\
&Differentiation (3/1)&2.064 (0.861 - 4.949)&0.104\\

\hline
&Cluster (Red/Black)&0.599 (0.251 - 1.429)&0.248\\
&Age at diagnosis&1.077 (1.041 - 1.115)&{\bf $<$0.001}\\
CEM-Dimension&Gender (male/female)&0.201 (0.671 - 2.228)&0.511\\
&Therapy (yes/no)&1.378 (0.491 - 3.871)&0.543\\
&Differentiation (2/1)&1.645 (0.722 - 3.750)&0.236\\
&Differentiation (3/1)&2.545 (1.108 - 5.846)&{\bf 0.028}\\
\hline

&Cluster (Red/Black)&0.574 (0.290 - 1.137)&0.111\\
&Age at diagnosis&1.076 (1.040 - 1.114)&{\bf $<$0.001}\\
CEM-Partial&Gender (male/female)&0.796 (0.691 - 2.284)&0.459\\
&Therapy (yes/no)&2.013 (0.975 - 4.156)&0.058\\
&Differentiation (2/1)&1.651  (0.737 - 3.696)&0.223\\
&Differentiation (3/1)&2.538 (1.102 - 5.841)&{\bf 0.029}\\
\hline

\end{tabular}
\end{table}

By setting a p-value threshold at $5\%$ in the Kaplan-Meier (KM) analyses, only the CEM algorithm did not stratify the stage I NSCLC data set into two clusters with different survival outcomes (Figure \ref{fig:KM}). However, notice that the KM analysis does not remove the effects of the covariates. Thus, we also analyzed using the Cox regression model, which considers the covariates. In this case, only CEM-Co identified a subgroup of stage I NSCLC with a statistically poorer survival outcome ($p<0.001$) (Table \ref{tab:cox}).

The reason why only CEM-Co was able to identify the poorer subgroup can be explained by how the covariates affect the clustering structure. 

\begin{figure}
\centering
\includegraphics[width=0.8\textwidth]{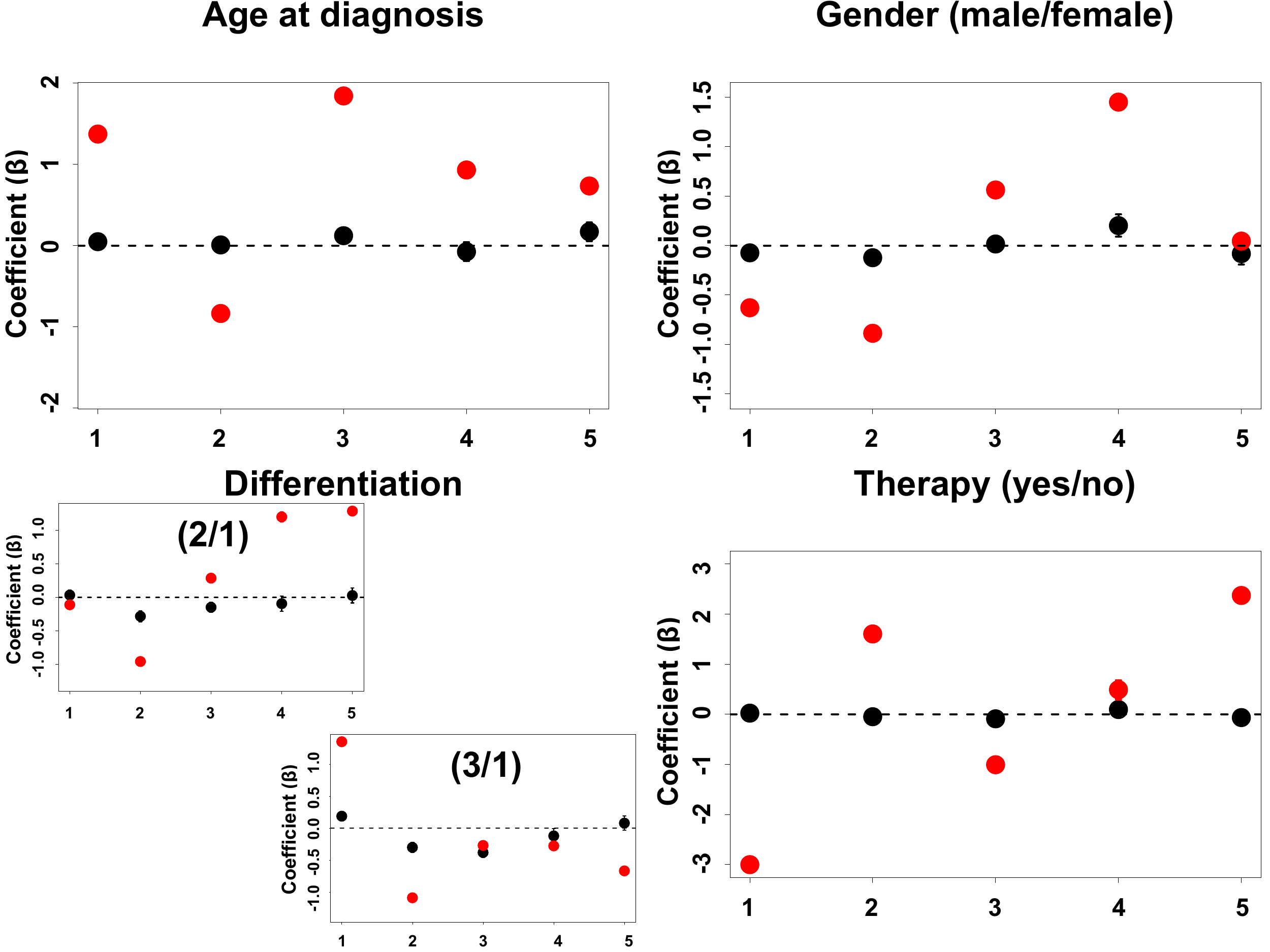} 
\caption{Covariates effects estimated by CEM-Co on stage I NSCLC. The $x$ and $y$-axes represent the $M=5$ dimensions of the items (principal components) and the coefficients ($\beta$) for each covariate effect, respectively. Red and black dots represent the patients with good and poor prognoses, respectively. Error bars represent the 95\% confidence intervals (some of them cannot be seen because they are too small). Except by therapy ($p=0.219$), all other covariates are significantly associated with the clusters' centroids (age at diagnosis - $p=0.002$, gender - $p=0.001$, and differentiation - $p=0.007$). Notice that the coefficients (covariates effects) are different between clusters (between red and black dots) and also among the five dimensions. The coefficients close to zero for the cluster with worse prognosis (black dots) can be explained probably by the small cluster size.}\label{fig:coef}
\end{figure}

 First, except by therapy ($p=0.219$), all other covariates significantly affect the clusters centroids: age at diagnosis ($p=0.002$), gender ($p=0.001$), and differentiation ($p=0.007$). This result confirms our hypothesis that covariates indeed affect the stage I NSCLC stratification. Second, the covariates affect both the clusters centroids and items' dimensions differently (Figure \ref{fig:coef}). However, both CEM-dimension and CEM-partial assume that the covariates affect equally the clusters centroids.

To verify the robustness of our results, we repeated all analyses by considering the number of clusters as $K=3$ (the second-best choice by BIC (Figure \ref{fig:BIC})). In this case, the largest cluster (red curve in Figure \ref{fig:KM}) was divided into two clusters (by all clustering methods). The Cox regression model for the clustering structure obtained by CEM-Co indicated that the smallest cluster presented a different survival outcome compared to the other two clusters, while the two largest clusters did not present different statistical survival outcomes between them. For the clusters obtained by CEM, CEM-dimension, and CEM-partial, the three clusters did not present statistically different survival outcomes among them. Therefore, the conclusions obtained for $K=3$ were the same at $K=2$.

Instead of selecting $M=5$ principal components (PCs), we also reanalyzed data considering four and six PCs. For four PCs ($M=4$, representing 46.41\% of the variance), the results are the same at $M=5$. Whether we consider six PCs ($M=6$, representing 55.8\% of the variance), CEM-Co clustered the data set into three groups, i.e., CEM-Co split the largest group into two groups. The smallest cluster presented a poorer survival outcome than the other two groups, while we did not observe any statistical difference between the two largest clusters. Therefore, the results seem to be robust even for different numbers of PCs.

Here we illustrated the importance of CEM-Co in stratifying a stage I NSCLC data set by minimizing the effects of age at diagnosis, gender, adjuvant therapy, and differentiation. However, we imagine that the flexibility of CEM-Co would make it applicable to other areas where clustering is a source of concern. For example, in large neuroimaging projects, it is known that the site where the data is collected strongly affects the results.

The EM algorithm has interesting theoretical properties as, for example, monotonicity and convergence to a stationary value. The proofs for our proposal are similar to the one presented in \cite{mclachlan2007algorithm}. One just need to condition on the observed value of ${\bf z}_i=({\bf z}_{1,i}$, ${\bf z}_{2,i}$, $\ldots, {\bf z}_{P,i})$.

One disadvantage of CEM-Co in the present form is that it only can be applied to data that can be modeled by a mixture of normal distributions. For example, we cannot apply CEM-Co to data sets represented by discrete variables. However, we think we can obtain a similar method for discrete data by using an appropriate probability distribution.

\section{Acknowledgments}
We would like to thank Alexandre Galv\~ao Patriota and Daniel Yasumasa Takahashi for reading the manuscript and providing very useful critics and comments.

\section{Funding}
CEMR was partially supported by S\~ao Paulo Research Foundation (FAPESP 2018/21934-5). AF was partially supported by FAPESP (2018/17996-5 and 2018/21934-5), CNPq (304876/2016-0), CAPES (Finance code 001), Alexander von Humboldt Foundation, and The Academy of Medical Sciences - Newton Fund.

\section{Conflict of interest statement}
The authors declare that the research was conducted in the absence of any commercial or financial relationships that could be construed as a potential conflict of interest.

\nocite{*}
\bibliographystyle{plain}
\bibliography{document.bib}
\addcontentsline{toc}{section}{References}

\end{document}